\documentclass[useAMS]{mn2e}
\usepackage{times}
\usepackage{epsfig}
\title[X-ray bump in GRB 121027A]{Variability of the giant X-ray bump in GRB 121027A and possible origin}
\author[Hou et al.]{Shu-Jin Hou$^{1,2,3}$, He Gao$^{4}$, Tong Liu$^{1,3,5}$\thanks{E-mail: tongliu@xmu.edu.cn}, Wei-Min Gu$^{1}$, Da-Bin Lin$^{6}$, Ya-Ping Li$^{1}$, \newauthor Yun-Peng Men$^{7}$, Xue-Feng Wu$^{2}$, Wei-Hua Lei$^{7}$, and Ju-Fu Lu$^{1}$\\
$^{1}$Department of Astronomy and Institute of Theoretical Physics and Astrophysics, Xiamen University, Xiamen, Fujian 361005, China\\
$^{2}$Purple Mountain Observatory, Chinese Academy of Sciences, Nanjing, Jiangsu 210008, China\\
$^{3}$Key Laboratory for the Structure and Evolution of Celestial Objects, Chinese Academy of Sciences, Kunming, Yunnan 650011, China\\
$^{4}$Department of Physics and Astronomy, University of Nevada Las Vegas, 4505 Maryland Parkway, Box 454002, Las Vegas, NV 89154-4002, USA\\
$^{5}$State Key Laboratory of Theoretical Physics, Institute of Theoretical Physics, Chinese Academy of Sciences, Beijing, 100190, China\\
$^{6}$Department of Physics and GXU-NAOC Center for Astrophysics and Space Sciences, Guangxi University, Nanning, Guangxi 530004, China\\
$^{7}$School of Physics, Huazhong University of Science and Technology, Wuhan, Hubei 430074, China}

\begin{document}

\pagerange{\pageref{firstpage}--\pageref{lastpage}} \pubyear{2014}

\maketitle

\label{firstpage}

\begin{abstract}
The particular giant X-ray bump of GRB 121027A triggered by \emph{Swift} is quite different from the typical X-ray flares in gamma-ray bursts. There exhibit four parts of the observed structural variabilities in the rise and decay phase of the bump. Considering the quality of four parts of the data, we can only analyze the data from about 5300 s to about 6100 s in the bump using the stepwise filter correlation method (Gao et al. 2012), and find that the $86^{+5.9}_{-9.4}~\rm s$ periodic oscillation may exist, which is confirmed by the Lomb-Scargle method (Scargle 1982). Furthermore, a jet precession model (Liu et al. 2010) is proposed to account for such a variability.
\end{abstract}


\section{INTRODUCTION}

Gamma-ray bursts (GRBs) are the most dramatic astronomical phenomena in the universe (Zhang \& M{\'e}sz{\'a}ros 2004; M{\'e}sz{\'a}ros 2006). The prompt emissions of GRBs last from a few milliseconds to thousands of seconds. The statistics of their durations are shown as a bimodal distribution (Kouveliotou et al. 1993), and therefore GRBs can be classified as short- and long-duration GRBs. The progenitors of them are believed to be the mergers of two compact objects (see e.g., Eichler et al. 1989; Paczy{\'n}ski 1991; Narayan et al. 1992) or the collapsars of massive stars (see e.g., Woosley 1993), respectively. Despite of the different progenitors, a rotating black hole (BH) surrounded by a neutrino-dominated accretion flow (NDAF, see e.g., Popham et al. 1999; Narayan et al. 2001; Gu et al. 2006; Liu et al. 2007, 2008, 2010a, 2010b, 2012a, 2012b, 2013; Sun et al. 2012; Li \& Liu 2013; Kawanaka et al. 2013; Xue et al. 2013) will be formed and therefore power GRBs via the neutrino annihilation or the BZ mechanism (Blandford \& Znajek 1977).

A tilted accretion disc surrounding a supermassive BH leads to the precession of the BH and results in an S- or Z-shaped jet as observed in galaxies (e.g., Lu \& Zhou 2005). The jet precession caused by the system of the BH and disc can also explain some periodic variabilities of X-ray binaries, such as SS 433 (Sarazin et al. 1980). Quasi-periodic feature observed by BATSE in the gamma-ray lightcurves motivates the idea that the GRB jet may be precessed (see, e.g., Blackman et al. 1996; Portegies Zwart et al. 1999). In the central engine of GRBs, either the misalignment of angular momenta of two compact objects or the anisotropic fall-back mass in collapsar may induce the precession between the BH and the accretion flow. According to the Bardeen-Petterson effect (Bardeen \& Petterson 1975), we suggest that in the jet precession model the BH can capture the inner part of the NDAF to conform with the direction of the angular momentum, and the outer part of the NDAF drives the BH and inner part to precess (Liu et al. 2010a; Liu \& Xue 2012). The model can be used to explain the temporal structure and spectral evolution of GRBs (Liu et al. 2010a), to simulate almost all types of the gamma-ray lightcurve of GRBs (Portegies Zwart et al. 1999; Lei et al. 2007), and to predict the intensities of the gravitational waves from GRBs (Romero et al. 2010; Sun et al. 2012). However, compared with the observations of the gamma-ray emission, there is little X-ray observational evidence of precession in GRBs.

In the paper, we will analyse the variability of the X-ray bump in GRB 121027A and discuss its possible origin by using our jet precession model. In Section 2, we describe the \emph{Swift}/XRT observations of GRB 121027A and use the stepwise filter correlation method, coupled with the Lomb-Scargle method, to present the analysis of the quasi-periodic X-ray lightcurve from about 5300 s to about 6100 s since trigger. In Section 3, the jet precession model is introduced. For the reasonable properties of the BH in the centre of collapsar, the model can explain the X-ray lightcurve of GRB 121027A. The conclusions and discussion are presented in Section 4.

\section{X-RAY DATA ANALYSIS OF GRB 121027A}

\begin{figure}
\centering
\includegraphics[width=0.5\textwidth]{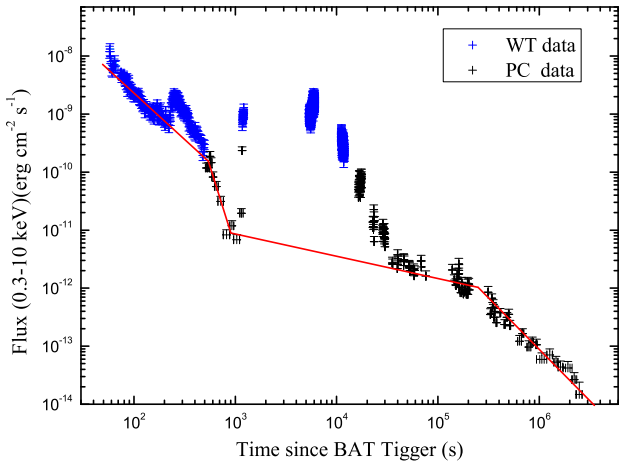}
\caption{{\em Swift}/XRT lightcurves of GRB 121027A. Blue data are the WT mode data, which also include the earlier WT settling mode data. The black data are the PC mode data. The red line is the fitting line. The light curve of XRT afterglow is consisted with six parts, including four power-law decay fragments, a flare and a giant bump. The giant bump is very rare in the X-ray lightcurves of GRBs.}
\end{figure}

The \emph{Swift}/XRT began observation at 67.4 s after GRB 121027A trigger and received about $\rm 2.4 \times 10^{6}$ s data (Evans et al. 2012). It is one of the typical ultra-long GRBs (e.g., Peng et al. 2013; Levan et al. 2014). The redshift was about 1.773 (Tanvir et al. 2012; Kruehler et al. 2012). Figure 1 shows the \emph{Swift}/XRT lightcurve, in which the data is downloaded from the UK Swift Science Data Centre at the University of Leicester (Evans et al. 2007, 2009). The X-ray lightcurve of GRB 121027A shows six components: a step decay phase from about 70 s to about 200 s with the temporal index about 1.8, a flare from about 200 s to about 500 s, another steep decay phase after the flare from about 500 s to about 1000 s with the power index about 7, a giant bump from about 1 ks to about 20 ks, a plateau phase from about 20 ks to about 100 ks (or from about 1 ks to about 100 ks) with the decay index about 0.34, and another normal decay phase from about 100 ks to the end with the temporal index about 1.44. It is rare that two continuous step decay phases with different temporal indexes exist in the afterglow.

Most of the lightcurve components are easily understood, except for the giant bump, which is the most interesting phenomenon in GRB 121027A observations (e.g., Peng et al. 2013; Wu et al. 2013). The related models were proposed, including the fall-back accretion process in collapsars (Wu et al. 2013) and a blue supergiant progenitor (e.g., Stratta et al. 2013). The shape of the particular X-ray bump of GRB 121027A is quite different from that of the typical X-ray flares in GRBs. Furthermore, we carefully analyze the substructures of lightcurve in the X-ray bump and find that there exist many violent oscillations. There are some substructures in the lightcurve from about 1170 s to about 1210 s, from about 5300s to about 6100 s, from about 11200 s to about 12000 s, and from about 16400 s to about 17600 s, as shown in Figure 2, which are quite different from the smooth lightcurves of the typical X-ray afterglow in GRBs. The red lines represent the links of the data in Figure 2 (a) and (d) and the smooth curves for the data in Figure 2 (b) and (c). Obviously, the timescales of oscillations in the four parts of the bump are from about tens of seconds to hundreds of seconds, which indicate that there may exist a rough time evolution in the lightcurve from about 1170 s to about 17600 s.

The stepwise filter correlation (SFC) method is used to decompose the variability components of lightcurves (see, e.g., Gao et al. 2012). The method can identify significant clustering structures of a lightcurve in the frequency domain, but it cannot give the statistical significance and location of the structure. It is based on a low-pass filter technique and progressively filter the high-frequency signals. Then it performs a correlation analysis between each adjunct pair of filtered lightcurves. The correlation coefficient as a function of the filter frequency would display a prominent ``valley'' feature around the frequency of the ``slow'' variability component, which can inspect the periodic signals. The detailed description can be found in Gao et al. (2012), which analyzed 266 GRBs observed by BATSE. Lei et al. (2013) applied the method to Sw J1644+57 and found a 2.7-day period.

We use the SFC method to process the data of the giant bump. However, the quality of the most data is not good. Only the data from about 5300 s to about 6100 s can be analyzed in details. For the observations from about 1170 s to about 1210 s and from about 16400 s to about 17600 s, the data are too rare to be analyzed by the SFC method. For the data from about 11200 s to about 12000 s, there is no indication of period by analysis using the method, because the modulation of flux and the fluence are too small.

\begin{figure*}
\centering
\includegraphics[width=0.45\textwidth]{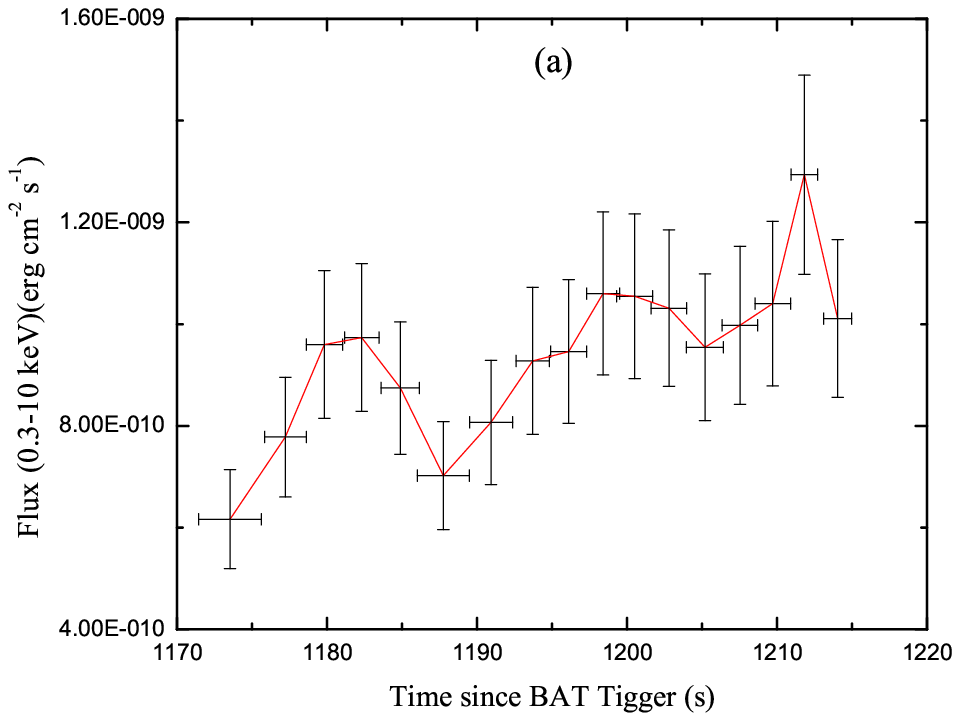}
\includegraphics[width=0.45\textwidth]{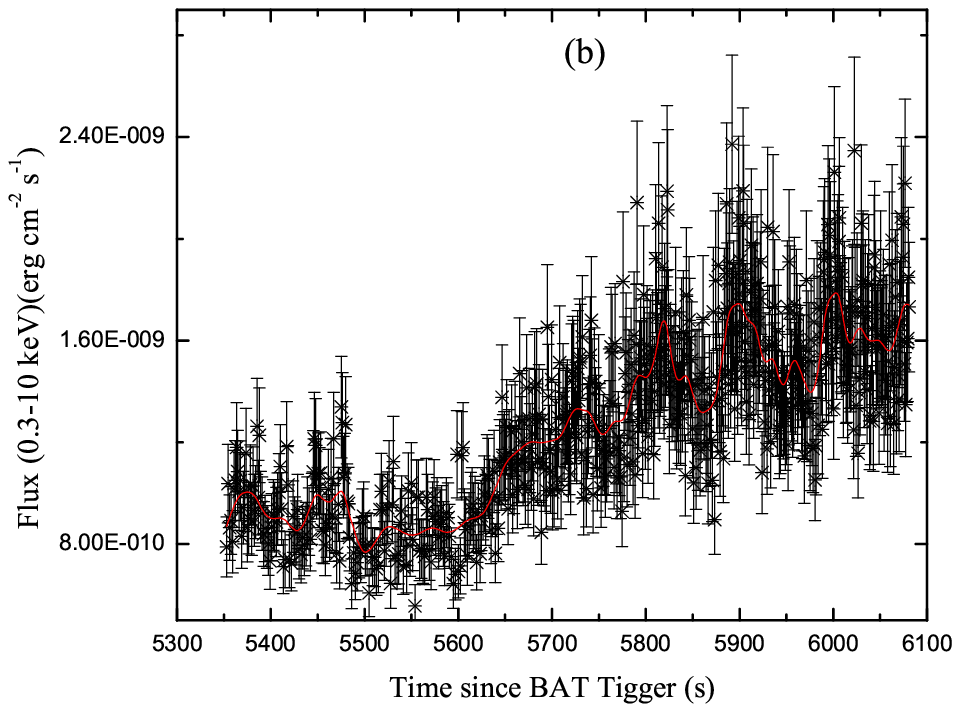}
\includegraphics[width=0.45\textwidth]{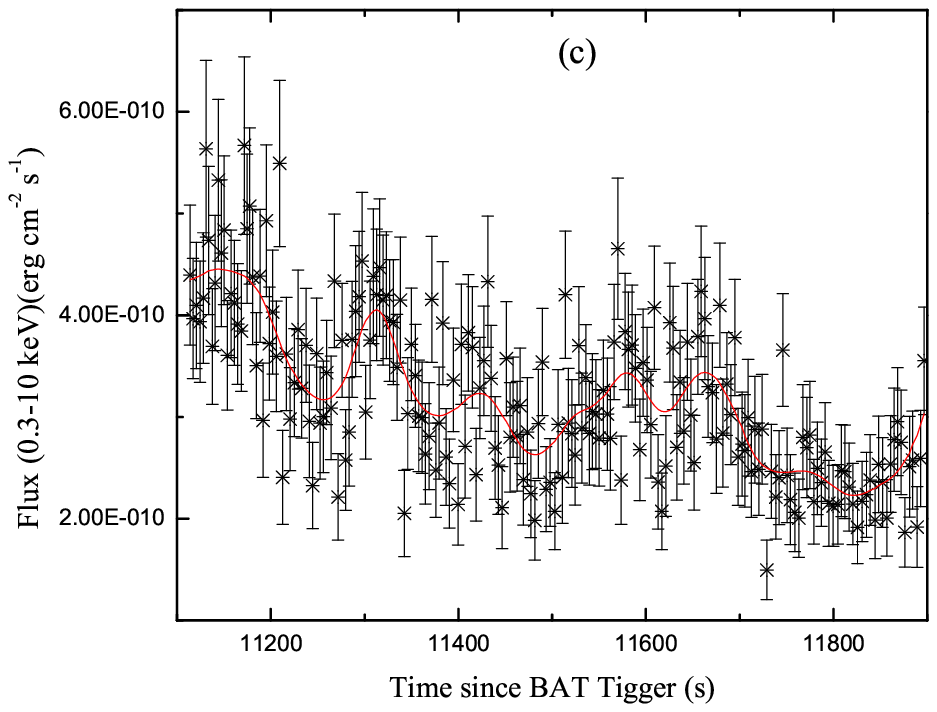}
\includegraphics[width=0.45\textwidth]{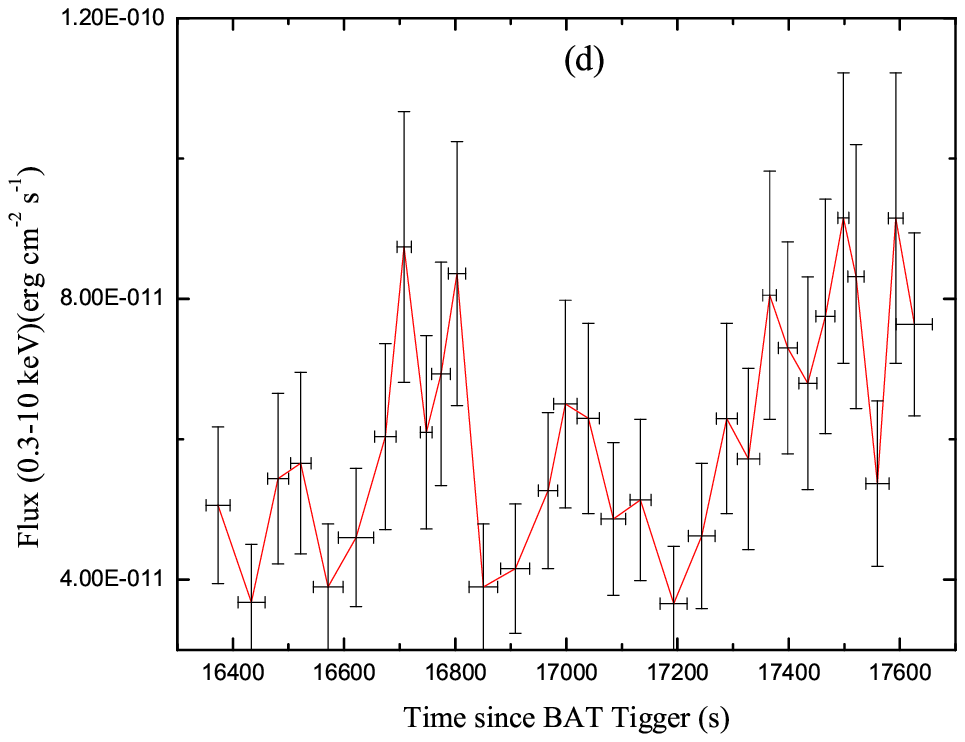}
\caption{Due to limit of the satellite orbit and observation mode conversion, there are only four fragments observed in the giant bump corresponding to four figures. It is obviously that the best profile is the data from about 5300 s to about 6100 s. The red lines represent the links of the data in (a) and (d) and the smooth curves for the data in (b) and (c).}
\end{figure*}

For the observations from about 5300 s to about 6100 s, because the data from \emph{Swift}/XRT observations are not equal to the bins, we first use the insert method to make the data equal to the interval for subsequent analysis. The errors of the data are taken as $5\%$ of the flux. We first perform an analysis to the lightcurve as shown in Figure 3(a). Using the method, the 86 s dip clearly shows up in Figure 3(b), which implies that there may exist a time structure lasting $\sim$ 86 s, corresponding to about 31 s in the rest frame. As shown in the lightcurve, we notice that the modulation of the flux after about 5700 s is very significant and there exist some peaks with the similar interval among each other, which indeed contains the periodic signals and dominates the analytic results. On the contrary, the circumstance of the data before about 5700 s is uncertain, because the small amplitude cannot be analyzed by SFC method separately.

Following Gao et al. (2012), we use Monte Carlo simulation to evaluate the error range of the period and quantify the significance for the detection of such quasi-periodic oscillations. First, for each time bin with a observed count rate $C$ and count rate error $\sigma_C$, which is taken as 5$\%$ of the flux, we can generate a mock count rate by randomly generating the data based on a normal distribution with ($C$, $\sigma_C$). 1000 mock light curves will be generated and we apply the SFC method to each mock light curve, and we can check whether the 86 s quasi-period exists. We propose that the percentage of mock light curves that contains 86s quasi-period, $c$, can be essentially acted as the significance parameter. It turns out $c=97.9\%$ for 5300 s to 6100 s data. Second, to estimate the period error, we perform the Monte Carlo simulation again but allow random seed error for each time to be $5\sigma$ of the original count rate error, i.e., generating the data based on a normal distribution with ($C$, $5\sigma_C$), and then identify the quasi-period, if exists. Figure 3(c) shows the distribution of quasi-period for mock light curves. The distribution fitting parameters with a statistical average of period and its variance are 84.24 s and 7.65 s ($1\sigma$ error). We thus propose to use $76.6~{\rm s} \sim 91.9~{\rm s}$ as the quasi-period error, i.e., the period is $86^{+5.9}_{-9.4}~\rm s$.

Furthermore, we use the Lomb-Scargle method (Scargle 1982) to examine the results by the SFC method, which is based on the discrete Fourier transform and is usually used to analyze the period of the exoplanet. As shown in Figure 3(d), if we set the range of the possible period from 80 s to 100 s and analyze the data from 5300 s to 6100 s by the method, a $\sim$ 86 s periodicity with the confidence of 97.3 $\%$ is obtained. Such a result is consistent with the analysis by the SFC method.

\section{JET PRECESSION MODEL AND APPLICATIONS TO GRB 121027A}

Although the progenitors of the two types of GRBs may be different, their central engines are similar, i.e., the BH hyperaccretion model. Such system drives an ultra-relativistic jet to produce both the prompt gamma-rays emission and afterglow in lower energy bands, whose orientation coincides with the direction of the angular momentum of the BH (e.g., Popham et al. 1999; Liu et al. 2007; Liu et al. 2010a; Liu \& Xue 2012). The different direction of the angular momentum between the BH and disc can induce the disc to tilt and jet to precess (Liu et al. 2010a; Sun et al. 2012).

\begin{figure*}
\centering
\includegraphics[width=0.43\textwidth]{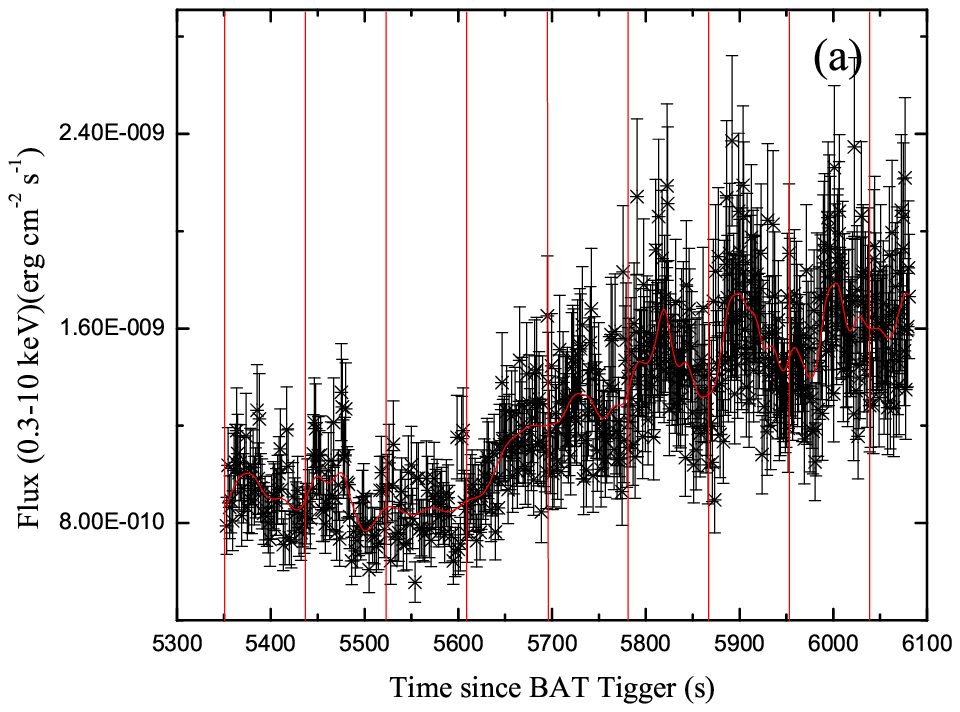}
\includegraphics[width=0.45\textwidth]{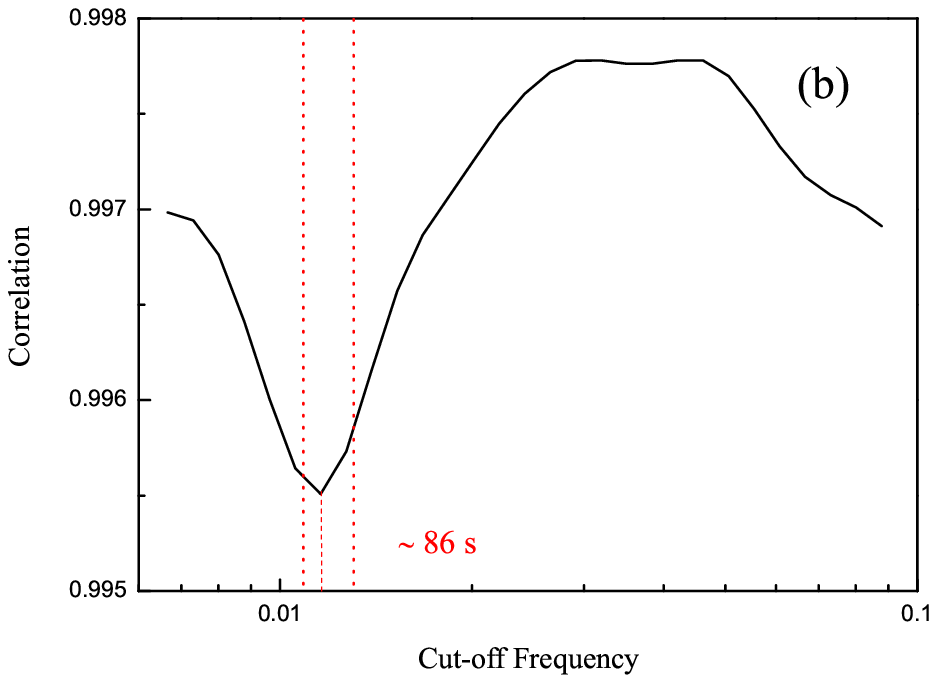}
\includegraphics[width=0.43\textwidth]{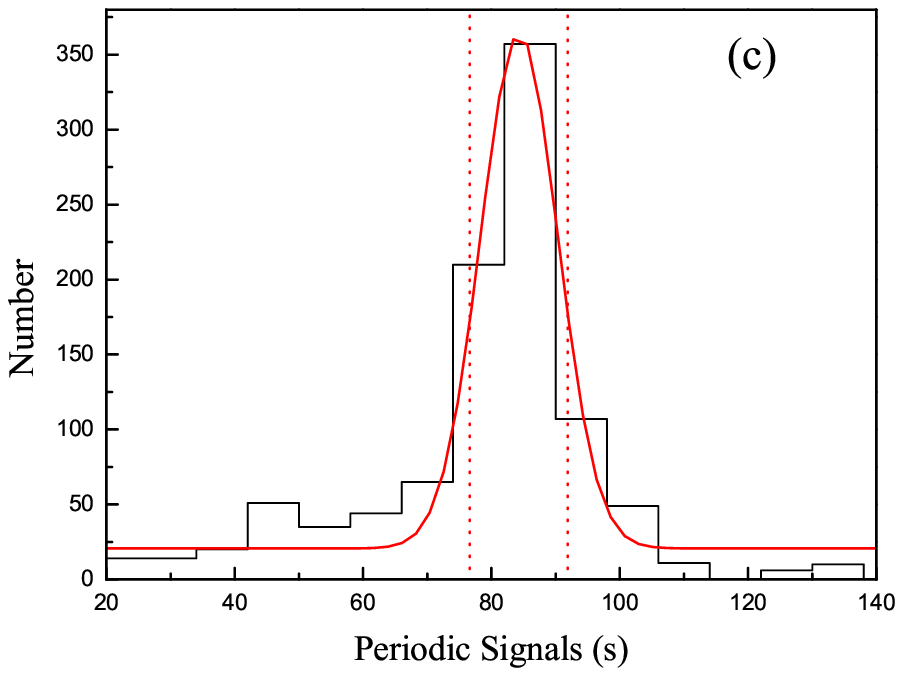}
\includegraphics[width=0.51\textwidth]{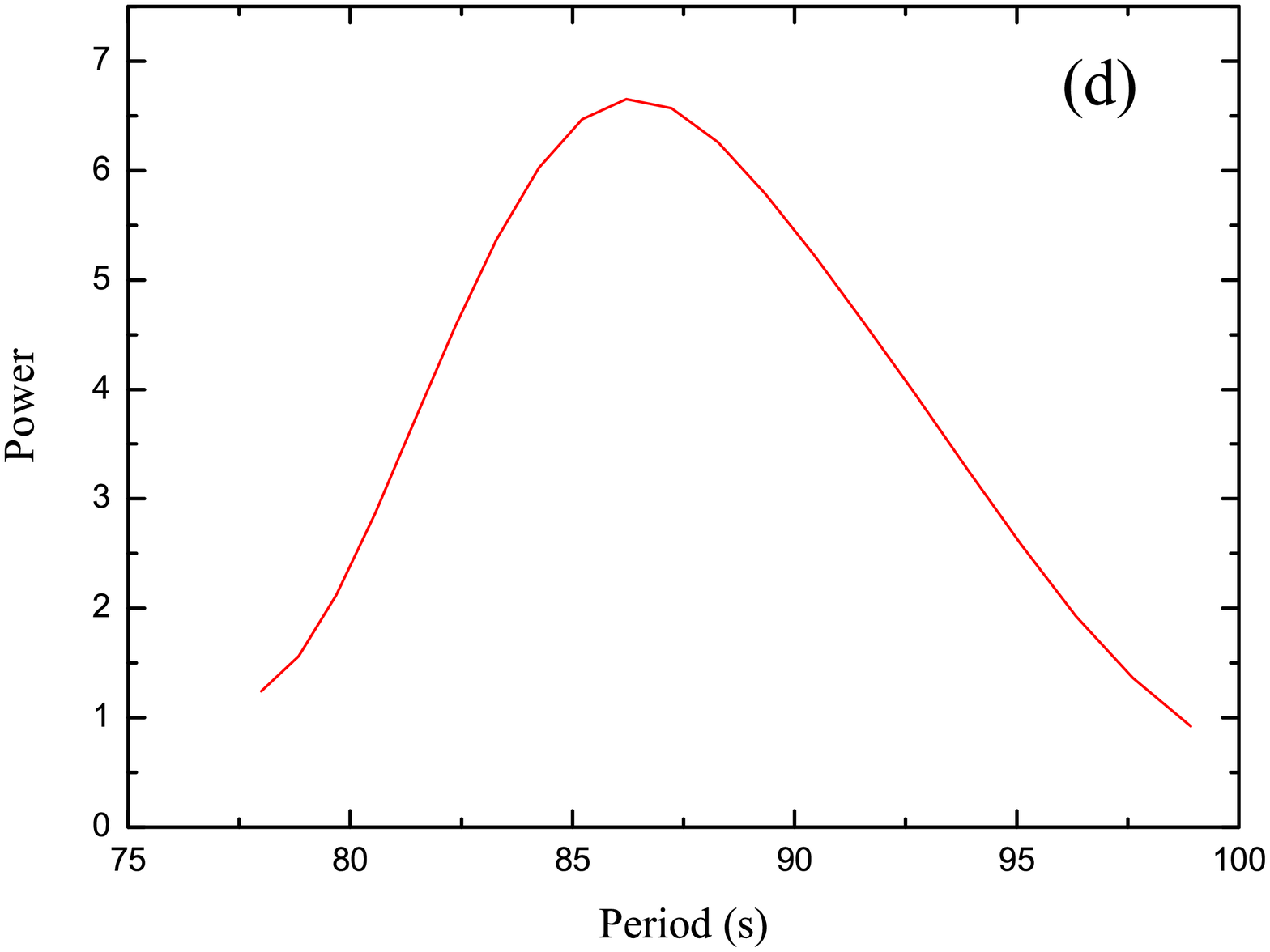}
\caption{Panel (a): the lightcurve of the giant bump from about 5300 s to about 6100 s. The red vertical lines show the duration $\sim$ 86 s. Panel (b): analysis of the data of the upper plane using the SFC method (Gao et al. 2012). The dip appears in 86 s, which implies that the period about 86 s exists. The red dotted lines show the error of the period. Panel (c): distribution of quasi-periodic signatures, which are driven from the Monte Carlo simulation method for the data from about 5300 s to about 6100 s. The red solid line is Gaussian fitting, which yields quasi-periodic signatures $84.24 \pm 7.65$ s (1 $\sigma$ error). The red dotted lines show the error of the period. Panel (d): analysis of the data of the upper plane using the Lomb-Scargle method (Scargle 1982).}
\end{figure*}

According to Equations (1), (2) and (5) of Sun et al. (2012) and Equations (5.6) and (5.7) of Popham et al. (1999), the analytic expression of the precession period $P$ can be expressed as (Hou et al. 2014)
\begin{equation}
P \approx 2793~ a_\ast^{17/13} (\frac{M}{M_\odot})^{7/13} (\frac{\dot{M}}{M_\odot~\rm s^{-1}})^{-30/13} \alpha^{36/13} ~\rm s,
\end{equation}
where $a_\ast$ and $M$ are respectively the dimensionless spin parameter ($0 <a_\ast <1$) and mass of the BH, $\dot{M}$ and $\alpha$ are respectively the accretion rate and the viscosity parameter of the disc.

If the X-ray bumps or flares originate from the BH hyperaccretion processes, we can connect the observed X-ray luminosity $L_{\rm X, iso}$ of the bumps or flares with the mass accretion rate $\dot{M}$ through
\begin{equation}
\dot{M} = \eta^{-1} c^{-2} L_{\rm X, iso},
\end{equation}
where $\eta$ includes the beaming effect and efficiency of converting accretion material to X-ray radiation. For the power mechanisms, both neutrino annihilations and magnetic processes have been included in the above equation.

For GRB 121027A, if the bump originates from the fall-back accretion process (e.g., Wu et al. 2013), the precession between the BH and disc may exist due to the possible anisotropic distribution of the angular momentum of the fall-back materials. From Equation (1), following the BH hyperaccretion process, the BH mass increases and the mass and angular momentum of the disc decreases, which may cause that the precession period increases with time. Thus the change of the BH accretion system can result in the evolution of the periods. Unfortunately, except the data from about 5300 s to about 6100 s, we cannot analyze the oscillations of the other data in the bump to estimate the evolution.

We can further test our jet precession model with the possible quasi-periodic variability of the bump. The average X-ray luminosity $L_{\rm X, iso}$ from about 5300 s to 6100 s is about $4.8 \times 10^{49}$ erg $\rm s^{-1}$ (Wu et al. 2013). We reasonably assume that $\eta$ is about $15\%$, so $\dot{M}$ is estimated by Equation (2) as nearly $1.8 \times 10^{-4}~M_\odot~\rm s^{-1}$. If we consider that the effect of the magnetic field in NDAF models, the low accretion rate can also ignite the disc to maintain neutrino emission (e.g., Kawanaka et al. 2013; Luo et al. 2013). In addition, for BZ mechanism to power GRBs, the requirements for accretion rate can be even lower.

Since the duration of the burst in the rest frame corresponds to the viscous timescale, which is inversely proportional to the viscosity parameter, a long-duration accretion process certainly requires a low viscosity parameter. For the accretion timescale lasting about ten thousands seconds, $\alpha$ can be estimated as $1.0 \times 10^{-4}$ by $\alpha \sim t_{\rm acc}^{-1}$ (e.g., Hou et al. 2014). The spin parameter $a_\ast$ is naturally assumed to be 0.9 after hyperaccretion process lasting thousands of seconds (e.g., Wu et al. 2013). The precession period is about $31^{+2.1}_{-3.4}~\rm s$ for GRB 121027A in the rest frame, the mass of the BH can be calculated as about $10^{+1.3}_{-2.0}~M_\odot$ with Equation (1). After thousands of seconds hyperaccretion process, the final mass of the BH is well consistent with the collapsar models (e.g., Popham et al. 1999), which indicates the jet precession model may naturally explain the origin of the bump in GRB 121027A.

\section{CONCLUSIONS AND DISCUSSION}

We roughly analyze the timescales of oscillations in the four parts of the bump, which indicate that there may exist the time evolution in the bump of GRB 121027A, and find there may exist $86^{+5.9}_{-9.4}~\rm s$ periodic signals from about 5300 s to about 6100 s by using the SFC method. The result is confirmed by the Lomb-Scargle method. If the bump originates from the fall-back accretion process, we argue that the quasi-periodic oscillations may be caused by the jet precession in the BH-NDAF system. Our model (Liu et al. 2010a) can explain the behaviours, and the final properties of the BH are well consistent with the collapsar models. Thus it is a possible method to test or estimate the mass of the BH in the centre of a GRB due to the quasi-period oscillations of the lightcurve in the future observations.

For the prompt emission, Gao et al. (2012) performed SFC method to 266 GRBs in BATSE sample. Although no quasi-periodic oscillations was claimed in their work, they indeed found that the majority of the bursts had clear evidence of containing a ``slow" variability component superposed on a rapidly varying time sequence. Furthermore, we searched all the \emph{Swift}/XRT samples. The light curves of most flares and afterglows are smooth, thus the quasi-periodic oscillations do not exist. For the ultra-long GRBs, we found a possible sample in the data of GRB 101225A. We used the SFC method to analyze its data from about 4950 s to about 7300 s, and did not find the periodic signals.

Furthermore, Fan et al. (2005) suggested that if jet powering the late X-ray flares is launched via magnetic processes, such as GRB 050724, the radiation of the flares is expected to be linearly polarized. As well as the bump of GRB 121027A, given the requirement for accretion rate ($\sim 1.8 \times 10^{-4}~M_\odot~\rm s^{-1}$) in the jet precession model, the jet is possibly dominated by the magnetic field. The bump including quasi-periodic signals may be one of the astronomical candidate sources of linearly polarized. Future GRBs observations by the POLAR detector (Bao et al. 2012) may test this possibility.

\section*{Acknowledgments}

We thank Prof. Yi-Zhong Fan  and Prof. Jun-Feng Wang for helpful discussions and the anonymous referee for very useful suggestions and comments. We acknowledge the use of the public data from the \emph{Swift} archives. Our work also made use of data supplied by the UK Swift Science Data Centre at the University of Leicester. This work is supported by the National Basic Research Program of China (973 Program) under grant 2014CB845800, the National Natural Science Foundation of China under grants 11103015, 11163003, 11222328, 11233006, 11333004, 11373002, and U1331101, the CAS Open Research Program of Key Laboratory for the Structure and Evolution of Celestial Objects under grant OP201305, and the Natural Science Foundation of Fujian Province of China under grant 2012J01026. X.F. Wu acknowledges support by the One-Hundred-Talents Program, the Youth Innovation Promotion Association, and the Strategic Priority Research Program ``The Emergence of Cosmological Structures'' of the Chinese Academy of Sciences (Grant No. XDB09000000).

\label{lastpage}

\clearpage

\end{document}